\documentclass[12pt]{article}

\usepackage{scicite}

\usepackage{times}

\usepackage[parfill]{parskip}    
\usepackage{graphicx}
\usepackage[all,line]{xy}
\usepackage{epstopdf}
\usepackage{booktabs}
\usepackage{mathtools}
\usepackage{stackrel}
\usepackage{caption}
\usepackage{subcaption}
\usepackage{mathtools}
\usepackage{url}
\usepackage{textcomp}

\usepackage{tabularx}

\usepackage{pdflscape}

\usepackage{multirow}

\usepackage{changes}

\usepackage{xcolor}

\newenvironment{sciabstract}{%
\begin{quote} \bf}
{\end{quote}}

\title{Evaluation of diagnostic test procedures for SARS-CoV-2 using latent class models:  comparison of antigen test kits and sampling for PCR testing based on Danish national data registries\\
{\small Short title: LCM evaluation of diagnostic tests for SARS-CoV-2}} 
\author{Jacob Stærk-Østergaard$^{1*}$, Carsten Kirkeby$^1$, Lasse Engbo Christiansen$^2$,  \\
Michael Asger Andersen$^3$, Camilla Holten Møller$^3$, Marianne Voldstedlund$^3$, \\
Matthew J. Denwood$^1$\\
\\
\normalsize{$^{1}$Department of Veterinary and Animal Sciences, University of Copenhagen,}\\
\normalsize{DK-1870 Frederiksberg C, Denmark,}\\
\normalsize{$^{2}$Department of Applied Mathematics and Computer Science, Technical University of Denmark,}\\
\normalsize{DK-2800 Lyngby, Denmark,}\\
\normalsize{$^{3}$Staten's Serum Institute,}\\
\normalsize{DK-2300 Copenhagen S, Denmark}\\ \\
\normalsize{$^\ast$Corresponding author; E-mail:  ostergaard@sund.ku.dk}
}
\date{ \today }

\begin{document} 


\baselineskip24pt


\maketitle


\begin{sciabstract}
Antigen test kits have been used extensively as a screening tool during the worldwide pandemic of coronavirus (SARS‑CoV‑2). While it is generally expected that taking samples for analysis with PCR testing gives more reliable results than using antigen test kits, the overall sensitivity and specificity of the two protocols in the field have not yet been estimated without assuming that the PCR test constitutes a gold standard. We use latent class models to estimate the \emph{in situ} performance of both PCR and antigen testing, using data from the Danish national registries. The results are based on 240,000 paired tests results sub-selected from the 55 million test results that were obtained in Denmark during the period from February 2021 until June 2021.

We found that the specificity of both tests is very high in our data sample ($>$99.7\%), while the sensitivity of PCR sampling was estimated to be 95.7\% (95\% CI: 92.8-98.4\%) and that of the antigen test kits used in Denmark over the study period was estimated at 53.8\% (95\% CI: 49.8-57.9\%). Our findings can be used as supplementary information for consideration when implementing serial testing strategies that employ a confirmatory PCR sample following a positive result from an antigen test kit, such as the policy used in Denmark. We note that while this strategy reduces the number of false positives associated with antigen test screening, it also increases the false negatives. We demonstrate that the balance of trading false positives for false negatives only favours the use of serial testing when the expected true prevalence is low. Our results contain substantial uncertainty in the estimates for sensitivity due to the relatively small number of positive test results over this period: validation of our findings in a population with higher prevalence would therefore be highly relevant for future work.
\end{sciabstract}

\paragraph*{Teaser description:} Estimating the diagnostic performance of antigen and PCR sampling as used routinely during the COVID-19 pandemic in Denmark.


\section*{Introduction}

Diagnostic testing procedures play a crucial role in the control of an infectious disease in terms of identifying infectious individuals and estimating the burden of infection. This is particularly true for diseases such as Severe Acute Respiratory Syndrome Coronavirus 2 (SARS‑CoV‑2), which spread rapidly due partly to asymptomatic individuals who have no other way of knowing that they are infected \cite{world2021recommendations}. However, it is essential to consider the diagnostic performance of each testing procedure, including potential errors related to imperfect sampling and accuracy of reporting results, when considering their utility within a disease control programme. This requires information on the real-world sensitivity and specificity of each of the available diagnostic test procedures.

There are several challenges involved with providing these estimates of sensitivity and specificity.  The first is that laboratory sensitivity and specificity as estimated under tightly controlled conditions may not be representative of the performance of the test in the field \cite{Jansen2019}, which also include challenges not directly related to the laboratory procedure such as contamination within the submitted sample. The second challenge is that the reference test against which we may wish to evaluate a new diagnostic test may itself be imperfect. This necessitates the use of latent class models (LCM) to analyse paired testing data in order to provide unbiased estimates of sensitivity and specificity for both tests being evaluated in the absence of a `gold standard' test. LCM based on the Hui-Walter paradigm were originally proposed over 40 years ago \cite{hui1980estimating} and have since become widely used for evaluating diagnostic tests within the {veterinary literature} \cite{bonde2010evaluation, rydevik2018evaluating, elsohaby2020accuracy}. Such tests also have great potential for further use within the human medical literature \cite{hartnack2021bayesian}. LCM have been used to estimate the sensitivity and specificity of reverse transcriptase polymerase chain reaction (RT-PCR), computed tomography (CT) and a number of other clinical and laboratory parameters for diagnosing SARS-CoV-2 infection \cite{symons2021statistical}, and to estimate the sensitivity and specificity of three commonly used tests for diagnosing COVID-19 \cite{hartnack2021bayesian}. However, the majority of studies that have evaluated antigen tests for SARS-CoV-2 to date have not used LCM methods to analyse the data, i.e. the studies have assumed that the reference test (typically RT-PCR) is perfect \cite{EC2021}. Despite this common assumption, all diagnostic tests are in fact imperfect, especially when considering extraneous sources of error such as mislabelling and contamination of the submitted sample. Therefore, using another imperfect test as a gold standard will bias the estimates of sensitivity and specificity of the test being evaluated.

There is one example of using LCM to assess the performance of antigen testing using laboratory samples \cite{saeed2021severe}. However, laboratory diagnostic test performance does not account for extraneous sources of error that are important in real-world settings, so data collected in the field provides a more relevant estimate of the performance of diagnostic tests from the perspective of disease control programmes. A previous study used field data to evaluate the diagnostic performance of antigen tests, but did so assuming that PCR sampling constituted a gold standard \cite{jakobsen2021detection}. To our knowledge, there are currently no studies that use LCM methods to provide unbiased estimates of test performance based on field data.

The scope of this paper is to apply LCM to estimate the \emph{in situ} sensitivity and specificity of PCR test sampling and antigen test kits performed during the SARS-CoV-2 epidemic in the Danish population between February and June 2021. We use the term `PCR sampling' to refer to the entire process between sample collection in the field, analysis of the sample using a Nucleic Acid Amplification Test (NAAT) procedure (which are predominantly RT-PCR in Denmark), and reporting of the result via the national database, to distinguish this concept from the RT-PCR test itself. Similarly, we use the term `antigen test' to mean the entire process including sampling and potentially imperfect use of the various different test kits that have been used in Denmark. Our analysis follows the STARD-BLCM reporting guidelines \cite{kostoulas2017stard}.


\section*{Results}

\paragraph*{Descriptive statistics}

\begin{figure}[ht]
	\centering
	\includegraphics[width=.9\textwidth]{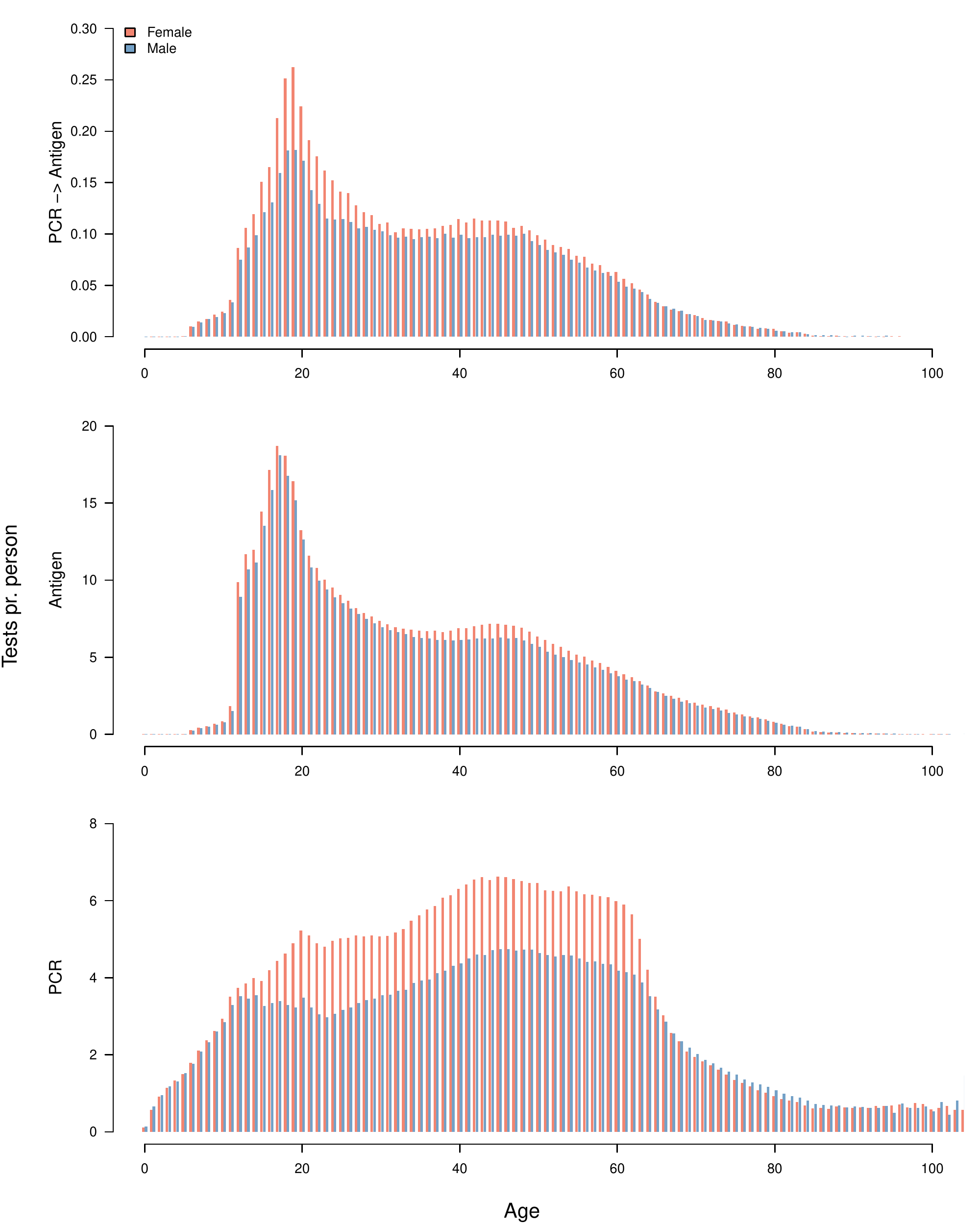}
	\caption{Representation of the study data relative to the Danish population. Data are at test level, which means that individuals can be represented several times. The y-axis reflects the average number of tests taken in the respective age group (males/females separately) during the study period from 1st February to 30th June. The higher rate of PCR sampling among females is partly due to screening for healthcare professionals, the majority of whom are female.}\label{fig: population}
\end{figure}

\begin{figure}[ht]
	\centering
	\includegraphics[width=.9\textwidth]{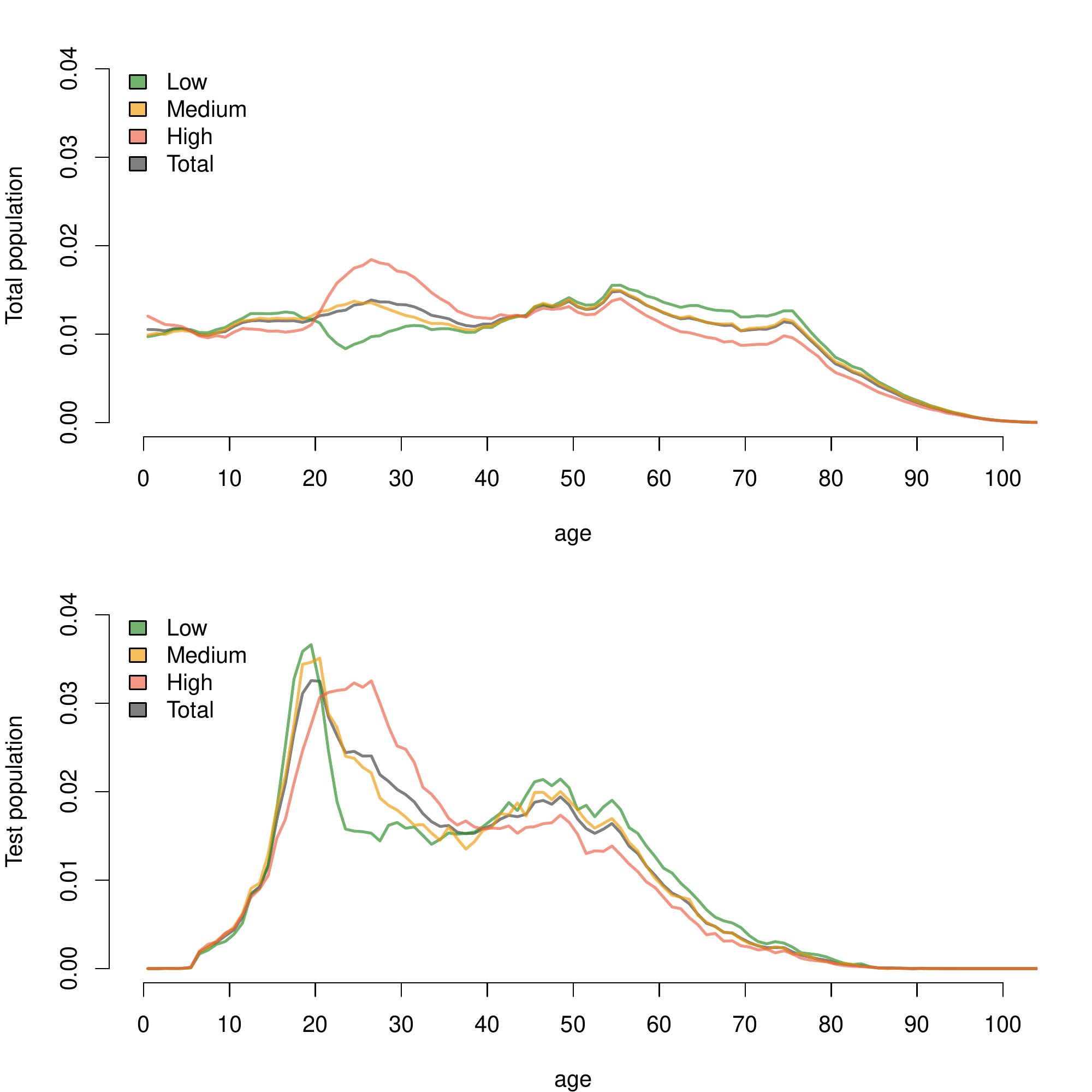}
	\caption{Age distribution across prevalence groups. The upper plot shows the age distribution among the full population in each group. The medium-prevalence group (yellow) follows the total population (black) to some extent. The lower plot shows the age distribution among the tests included in the study. It is skewed towards younger individuals. This is partially due to the heavy use of antigen testing in primary schools, high schools and universities.}\label{fig: group demographics}
\end{figure}

A total of 239,221 test pairs, where an antigen test kit was taken within 10 hours of being sampled for PCR testing, were available from 222,805 individuals. Among these, 77,439 pairs were from people living in high-prevalence parishes, 62,837 were from medium-prevalence parishes and 57,992 from low-prevalence parishes. The vaccinated group had 40,953 test pairs. We assume that each test pair was performed on separate samples for antigen and PCR testing from the same individual, however a small number of sample pairs (n=360) were registered with exactly the same date/time stamp, so it is possible that some of these tests were performed on the same sample. The tallies for each combination of test results is presented in Table \ref{tab: RT-PCR -> antigen}, while Figure \ref{fig: group demographics} shows the age distributions of the high/medium/low-prevalence groups. The medium-prevalence group shows similar age characteristics to the full population, whereas the high/low-prevalence groups have an over/under-representation of 20-40-year-old individuals and an under/over-representation of $>$55-year-old individuals, respectively, which aligns well with 20-40 year olds being the main source of positive tests during May-June 2021. The demographic makeup of individuals taking the test combination PCR$\to$antigen generally follows that of the population taking antigen tests, with some additional sex bias (Figure \ref{fig: population}). Overall, the frequency of antigen testing has shifted towards younger individuals for both males and females, possibly reflecting that antigen tests have been used in Denmark to screen for SARS-CoV-2 in primary schools, high schools and university campuses. The demography of the population undertaking PCR sampling is more evenly distributed across ages, with a higher rate for females than males. The higher rate of female PCR sampling might be the cause for the higher rate of females in the PCR$\to$antigen data. The higher rate of female PCR sampling is partially due to this procedure being used to screen healthcare professionals, the majority of whom are female. 

\begin{figure}[ht]
	\centering
	\includegraphics[width=.7\textwidth]{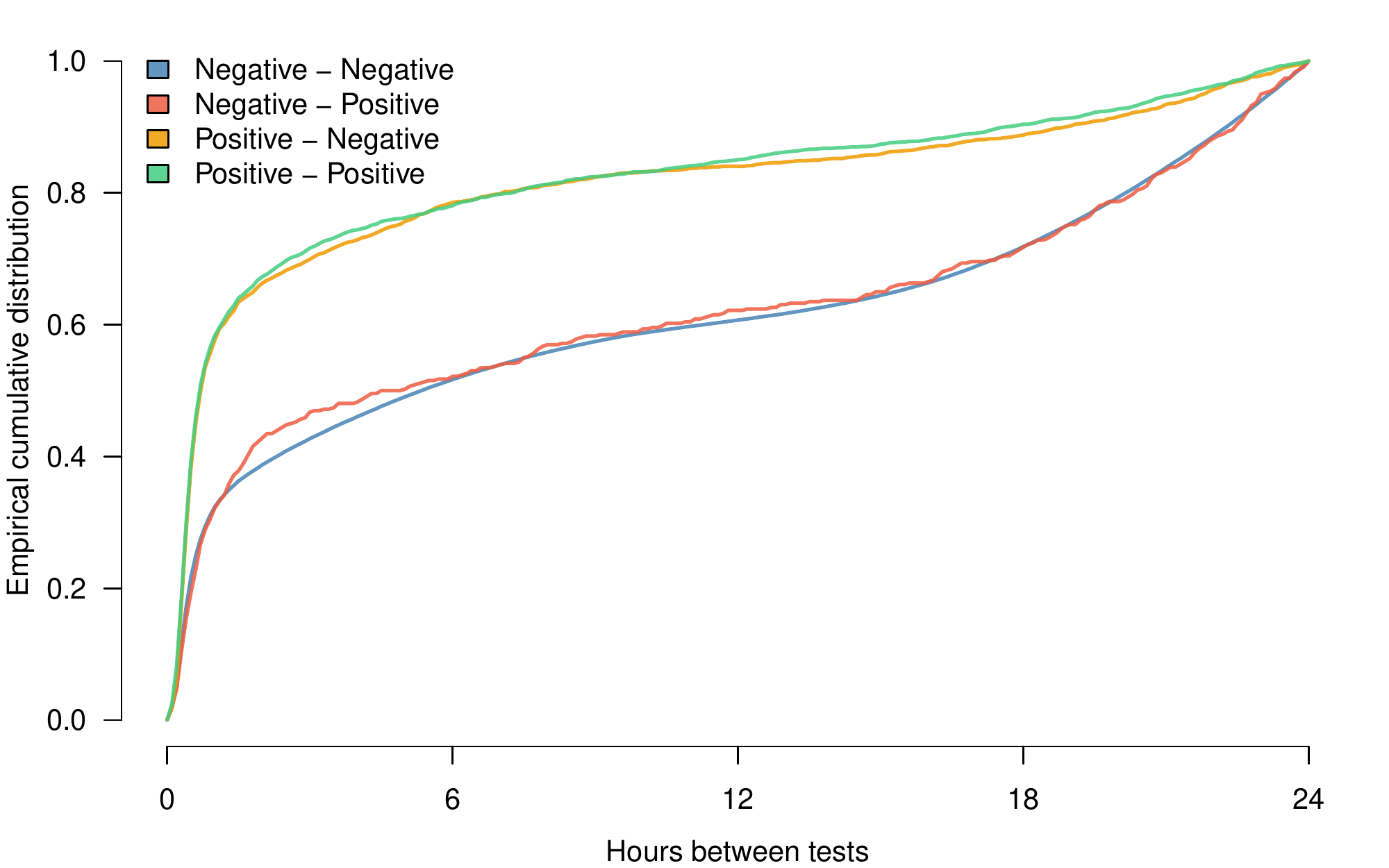}
	\caption{Distributions of time intervals between PCR and antigen tests (in hours). The legend describes the test outcomes (PCR – Antigen). Note that a positive result from a PCR sample is followed more rapidly by an antigen test than is the case for a negative result from a PCR sample.}\label{fig: waiting times}
\end{figure}

The distribution of time intervals between a positive PCR sample and a subsequent antigen test does not depend on the result of the antigen test (Figure \ref{fig: waiting times}). There was a slightly longer interval on average between a negative PCR sample and a negative antigen test (negative $\to$ negative) compared to that between a negative PCR sample and a positive antigen test (negative $\to$ positive). However, a Kolmogorov-Smirnov (KS) test did not show a statistically significant difference ($p$-value$=0.370$). 


There is, however, a substantial reduction in time intervals following a positive PCR sample compared to that following a negative PCR sample, regardless of the antigen test result (Figure \ref{fig: waiting times}). This difference is statistically significant as measured by the KS test ($p$-value $<0.001$ for any combination). This indicates that individuals who are more likely to have a positive PCR sample (perhaps because they are a high-risk secondary contact and/or because they have clinical symptoms) are more likely to take an antigen test within a shorter period of time following their PCR sample.


\paragraph*{Statistical modelling}

The frequencies for the four combinations of paired test results for each of the three populations are shown in Table \ref{tab: RT-PCR -> antigen}. The Hui-Walter model that was fit to these cross-tabulations converged and produced effective sample sizes above 1,000 for all parameters. Results are presented in Table \ref{tab: HW estimates RT-PCR -> antigen}. The specificity of PCR sampling was estimated to be 99.85\% (95\% CI: [99.73;99.97]), and the specificity of antigen testing was estimated to be 99.93\% (95\% CI: [99.91;99.96]). The corresponding sensitivity estimates were 95.68\% (95\% CI: [92.79;98.43]) for PCR sanmpling and 53.82\% (95\% CI: [49.83;57.93]) for antigen testing for unvaccinated individuals. The corresponding sensitivity estimates for vaccinated individuals were 97.44\% (95\% CI: [91.55;100]) for PCR sampling and 56.01\% (95\% CI: [44.50;69.84]) for antigen testing. Based on the overlapping confidence intervals, there is no evidence of a true difference in sensitivity between vaccinated and unvaccinated individuals for either test.

The sensitivity and specificity estimates are the sole focus of this study, however, the prevalence estimates corresponding to each of the groups are presented in the appendix (Table \ref{tab: HW prevalence RT-PCR -> antigen}). The prevalence estimates for the three unvaccinated groups are highest for the high-prevalence group and lowest for the low-prevalence group, as would be expected given the way in which these groups were artificially constructed. The confidence interval limits of the low-prevalence unvaccinated group overlaps with that of the vaccinated group, indicating a somewhat similar prevalence between these two groups. 

Results of the sensitivity analysis, varying the maximum allowable time between PCR sampling and antigen testing (time lag), are given in appendix \ref{app: additional results}. The sensitivity of antigen tests was estimated consistently for all analyses, with 95\% CI substantially overlapping across the range of time lag values. For PCR sampling, a small decrease in the median estimate of sensitivity was observed from a maximum of 97.65\% at 2 hours to a minimum of 94.66\% at 24 hours, however the 95\% CI produced for each time lag value overlapped substantially. Similarly, the specificity of both procedures was estimated consistently across varying time lag values.



\paragraph*{Implications for serial testing}

For serial testing, the estimated overall specificity is 100\%, while the sensitivity is 51.48\% (95\% CI: [47.37;55.96]), see Table \ref{tab: HW estimates RT-PCR -> antigen}.

During the study period, there were 35,530 positive antigen samples in Denmark from a total of 32,789,084 tests, and 109,922 positive PCR samples from a total of 22,052,829 samples. Adjusting the observed proportion of positive results from each test using the Rogan-Gladen estimator \cite{Rogan1978}, we found a corrected prevalence of 0.0916\% based on antigen tests and 0.4121\% based on PCR sampling. This discrepancy can be explained to some extent by the fact that while antigen tests are being used for screening in the general population, PCR sampling is used for diagnostic purposes, i.e. for confirmation, near-contacts and screening in hospitals. Therefore, PCR sampling might be more often applied in sub-groups where the prevalence would be expected to be higher. Of the 35,530 positive antigen tests, 28,366 were followed by a confirmatory PCR sample within 3 days, corresponding to around 80\%. Of these follow-up PCR samples, 11,985 were negative, thus releasing these individuals from quarantine. We therefore adopted 80\% as a reference of how many positive antigen tests would be followed (and potentially superseded) by a PCR sample in the scenarios of varying prevalence.

Table \ref{tab: estimate fp and fn cases for antigen and serial testing} shows the estimated total number of test result cases (and 95\% confidence limits) of false positives and false negatives for antigen testing alone compared with serial testing. Evidently, the number of false positive cases at national level is quite stable around 21-22,000 with prevalence between 0.01\% and 4\%. As Table \ref{tab: estimate fp and fn cases for antigen and serial testing} shows, the serial testing scheme effectively removes almost all false positive cases. However, as the prevalence increases, so does the number of false negatives. Since the sensitivity of serial testing is lower than that of antigen testing alone, the rate of increase in false negative cases for serial testing is higher than for antigen tests. Table \ref{tab: fp fn difference estimates} presents the estimated increase in false negative cases when changing from antigen testing alone to serial testing, as well as the decrease in false positives and the balance between these two. As shown, the balance is in favour of the serial testing scheme when the prevalence is low, since the number of false positives that are eliminated exceeds the expected increase in false negatives. However, a higher prevalence of $\approx3\%$ favours antigen testing alone, since the median number of false negatives outweighs the false positives in this scenario. At the lower limit of the confidence interval (2.5\%), the balance tips at a prevalence of $\approx1\%$. These results show that the implicit trade-off between sensitivity and specificity in serial testing should be taken into account if this strategy is used during a disease outbreak.


\section*{Discussion}

To our knowledge, this study represents the first use of LCM to assess the overall diagnostic utility of sampling for PCR testing and antigen kit testing for COVID-19 in the field. The study period in this paper covers the 5 months from 1st February to 30th June 2021, with data sub-sampled from the complete database of Danish test results. The pandemic in Denmark peaked at the end of 2020, before the mass administration of vaccines began in early 2021. Antigen tests were rolled out during 2021, with daily tests beginning to increase rapidly by February. As such, the study period covers a period of increasing test numbers and an initially low incidence that increased during the study period and peaked around the end of May. During this period, vaccines were also administered, beginning with the oldest age groups and others with a high risk of hospitalisation, continuing with younger and less vulnerable groups as the study period went on. From March 2021, there was a transition from national lockdown to a re-opening of schools and shops, with social activities permitted once again. This indicates that the study period covers a time during the pandemic when multiple factors influenced the incidence rates. A main factor in this study was the shift to younger generations being the main driver of the continued infections. The Hui-Walter model paradigm requires the use of multiple populations with differing prevalence but identical test specificity and sensitivity. In order to maximise the ability of the model to extract information from the data, we used artificial stratification based on the expected prevalence in the parish of residence. 

We found that the specificity for both test procedures was estimated to be close to 100\%: 99.85\% (95\% CI: [99.73;99.97]) and 99.93\% (95\% CI: [99.91;99.96]) for PCR sampling and antigen testing, respectively. These estimates are in line with the clinical performance of the antigen test kits used in Denmark (see Table \ref{tab: antigen clinical performance}), and our 95\% confidence intervals represent relatively precise results due to the high number of true negative individuals. The sensitivity was estimated to be 95.68\% (95\% CI: [92.79;98.43]) for PCR sampling and 53.82\% (95\% CI: [49.83;57.93]) for antigen testing. These estimates are more uncertain that those of specificity due to the relatively low number of true positive individuals in Denmark over this time period. Our model allowed the sensitivity of the group of vaccinated individuals to differ from that of unvaccinated individuals. However, the 95\% confidence intervals for these estimates overlapped substantially, and we therefore conclude that there is no evidence that the performance of these diagnostic procedures is dependent on the vaccination status of the individual being tested. However, these findings should be considered to be uncertain due to the relatively low number of vaccinated individuals in our study and correspondingly wide 95\% CI for the sensitivity estimate particularly in vaccinated individuals.

Despite these limitations, our results show that the sensitivity of PCR sampling in Denmark over our study period was relatively high (i.e. over 91.5\% in the worst case, and potentially as high as 100\% in vaccinated individuals). However, it is important to emphasise that it should not be considered a gold standard when evaluating the performance of antigen testing, since this approach will lead to a downward bias in the estimated performance of the antigen test. The imperfect nature of PCR sampling also affects the use of confirmatory PCR testing following a positive antigen test. This serial testing scheme has been employed in Denmark to reduce the number of false positives generated by the routine use of antigen tests. However, our study highlights the cost of this strategy in terms of increased false negative cases. Indeed, our findings suggest that the expected number of false negatives have increased during the study period due to the sensitivity and specificity of the serial testing scheme. As we also demonstrate, this depends heavily on the true prevalence, with the reduction in false positives expected to be equal to the increase in false negatives at a prevalence of around 3\%. As such, the serial testing strategy is justifiable when the prevalence is low, but as infection rates increase, decision makers must consider whether a trade-off of 1:1 is acceptable. Given a confirmatory PCR sample follow-up rate of 80\% combined with the estimates in Table \ref{tab: fp fn difference estimates} and an assumed true prevalence between 0.1\% and 0.4\%, the increase in false negatives is expected to lie somewhere between 610 ($=0.8\cdot763$) and 2,443 ($=0.8\cdot 3054$) cases, while the corresponding reduction in false positives is expected to be between 17,645 ($=0.8\cdot22,056$) and 17,592 ($=0.8\cdot21,990$). This implies a trade-off between 29:1 to 7:1 in favor of reducing false positives. 

Compared to the results of Jakobsen et al (\cite{jakobsen2021detection}), we found a similar, although marginally higher, specificity for antigen testing. The slight increase in the estimate for specificity is most likely due to false negative PCR sample results being erroneously attributed to false positive antigen tests. However, our estimate for sensitivity (54.77\%) is substantially lower than the value of 68.9\% that was previously reported. There are multiple possible reasons for this discrepancy. Firstly, data in Jakobsen et al (\cite{jakobsen2021detection}) were collected under a research protocol and therefore under more tightly controlled conditions than would be expected in the field, which would be expected to increase the diagnostic test performance. Furthermore, the previous study used data collected on 26th December 2020, and out of the 4,697 sampled individuals, 705 (15\%) reported symptoms, while 3,008 (64\%) reported no symptoms. For the symptomatic group, the sensitivity climbed to 78.8\%, while for the group without symptoms, the number was 49.2\%. Based on a voluntary questionnaire when booking a time for PCR sampling in Denmark, 10.0\% reported that they booked a PCR test ``due to showing COVID-19 symptoms". From February 2021 to March 2021, the group of self-reported symptoms ranged from 4.7\% to 7.2\%. As such, the proportion of individuals with symptomatic disease in the real world dataset is substantially lower than that for the previous study, which may be expected to negatively impact the overall sensitivity of antigen tests.

As with all LCM, we must consider the implicit meaning of the latent class that we are estimating \cite{kostoulas2017stard}. The definition of this latent class is tied to the statistical concepts inherent to the LCM, and represents the underlying `true state' on which the test results can be considered to be conditionally independent \cite{Toft2005}. However, we note that the `true state' in the LCM sense may not perfectly match the biological definition of `infected' or even `infectious'. This is because RT-PCR tests detect viral RNA, while antigen tests detect viral antigens. As such, the latent state implicitly defined by the LCM is `presence of viral RNA and antigens in the samples' rather than `individual is infected with virus'. It is therefore possible that part of the reason for the estimated imperfect sensitivity of the PCR sampling as estimated by the LCM is due to detection of either early-stage infection or late-stage infection corresponding to detectable levels of viral RNA but absence of viral antigens, which may be considered by the LCM as a `true negative'. In addition to this, it is also important to take into account the self-selection bias caused by non-random sampling of individuals for testing. It is relatively uncommon for an individual to take an antigen test within 10 hours following a PCR sample, and we cannot reasonably expect that these individuals are representative of the general population. The true interpretation of the prevalence estimates presented here is therefore: the average prevalence of virus shedding in each of the subgroups among the individuals who chose to take a PCR sample with a follow-up antigen test within 10 hours over the 5-month period. There is also a strong temporal confounding with these estimates due to the gradual roll-out of vaccines in Denmark – the vaccinated group is predominantly represented by tests taken later in the time series when the prevalence can be expected to be lower. It may also be tempting to compare the prevalence estimate from the unvaccinated groups to that of the vaccinated groups. However, there is substantial temporal bias in terms of the proportion of individuals vaccinated over this time period \cite{SSIdashboard}, so vaccination status is therefore confounded with the underlying temporal trends of disease burden in the general population.  Furthermore, there was variation in the official policy towards routine testing between vaccinated and unvaccinated individuals. We therefore note that the prevalence estimates for each of the four groups should not be interpreted as the prevalence of either clinical disease or SARS-CoV-2 infection in these groups. These are provided purely for the context of the LCM and should not be interpreted as being representative of any unbiased prevalence estimate that could be made over this time period. However, although we do not consider the prevalence estimates to be directly useful, they are necessary parameters within the Hui-Walter framework and we report them as advised by the STARD-BLCM reporting guidelines \cite{kostoulas2017stard}.

There are a number of assumptions and limitations associated with this study. The data were collected in such a way that test pairs were used when an antigen test was taken within 10 hours of a PCR sample. Since the usual response time for the PCR sample is between 10-36 hours, with a mean of 14 hours, the PCR result would not have been known before taking the antigen test in almost all cases. The antigen test can therefore be assumed to be independent of the PCR test, conditional on the underlying latent disease state of the individual. It is possible that a small number of individuals may have known their PCR sample result before having an antigen test, which may have affected their decision to take an antigen test. However, we believe that this is unlikely and therefore does not have a strong impact on our conclusions. Our sensitivity analysis, in which we alter the time period between tests produces qualitatively similar results, which supports this conclusion. The second important assumption made by our analysis is that the individuals included in the LCM analysis are representative in terms of the expected test sensitivity and specificity. In our case, this means that we should have no reason to suspect either a higher or lower sensitivity or specificity for individuals having both tests within 10 hours compared to individuals who have only a single test. In reality, it may be the case that our model data include a higher proportion of individuals with clinical disease than is true of the general population: it is therefore possible that we overestimate the sensitivity of both tests to some extent. However, we can think of no reason that the specificity estimates may be in any way biased by our data selection criteria. It is important to note that we expected the prevalence estimates to be heavily biased because we expect individuals who take both tests to have a higher than average probability of testing positive. This bias is also borne out by our results, which show far higher prevalence estimates than are believed to be the case for Denmark. However, this bias in prevalence estimates does not impact our study because estimation of prevalence is not our aim: the only important assumption is that the estimates of sensitivity and specificity are unbiased. It is also important to recognise that our estimates are based on a data sample taken from Denmark over the period February to July 2021:  findings may differ in future studies based on different datasets, particularly if fundamental properties of test procedures differ over time. Finally, we emphasise that our results refer to overall sensitivity and specificity in the field, which includes potential sources of error that are extraneous to the tests themselves such as sample contamination, mislabelling and misreporting of results. These estimates of operational sensitivity and specificity are highly relevant when evaluating diagnostic testing in terms of the overall effectiveness within a disease control programme.

\section*{Conclusions}

Our results show that the overall sensitivity of antigen testing and PCR testing was around 54\% and 96\%, respectively, when used as part of the Danish national control programme for SARS-CoV-2 between February and July 2021. However, our estimates for sensitivity are relatively uncertain due to the low number of true positive individuals in our dataset - validation of these findings in a population with higher prevalence would therefore be valuable. We also found that the overall specificity was close to 100\% for both procedures, and that the use of confirmatory testing based on PCR sampling following positive antigen tests increase the number of overall false negative results. When the prevalence is low (\textless1\%), a small increase in false negatives may be tolerated due to the relatively large decrease in false positives, but when the prevalence is high (\textgreater3\%) the increase in false negatives exceeds the decrease in false positives. The imperfect performance of PCR sampling in the field should therefore be accounted for when considering COVID-19 testing policies.

\section*{Materials and methods}

The centralised nature of record keeping within the Danish healthcare system provides a natural way in which to evaluate diagnostic test performance in the field. All medical procedures in Denmark are recorded on a centralised database system  against an individual's personal identification number in such a way that national-level analysis of the results is possible. Similarly, all Danish SARS-CoV-2 test results are recorded in the Danish Microbiology Database (MiBa) \cite{Schonning2021}. PCR sampling is the recommended procedure for diagnostic testing and contact tracing in Denmark, and these tests have been used extensively throughout the SARS-CoV-2 epidemic to survey and control the spread of the infection. Antigen tests have also been offered as an additional testing option since the beginning of 2021, and these have mainly been used as a screening test in schools, workplaces and to support the implementation of the COVID-19 passport\footnote{Corona passports were introduced in Denmark in March 2021 following the re-opening from a national state of lockdown. A valid passport (following a negative test result, vaccination or immunity from previous infection) could be used for access to various social activities as well as physical attendance at many workplaces.}.

Information about the precise date and time of each PCR sample and antigen test, as well as the result of the tests and identity of the individual being tested, can be extracted from the Danish national registries. This makes it possible to identify individuals who had both PCR and antigen tests within a short time window, so that the relevant test results can be used to provide a large-scale evaluation of the in-situ performance of PCR sampling and antigen testing within the Danish population. 


\paragraph*{Data sources}

\begin{figure}[ht]
	\centering
	\includegraphics[width=.7\textwidth]{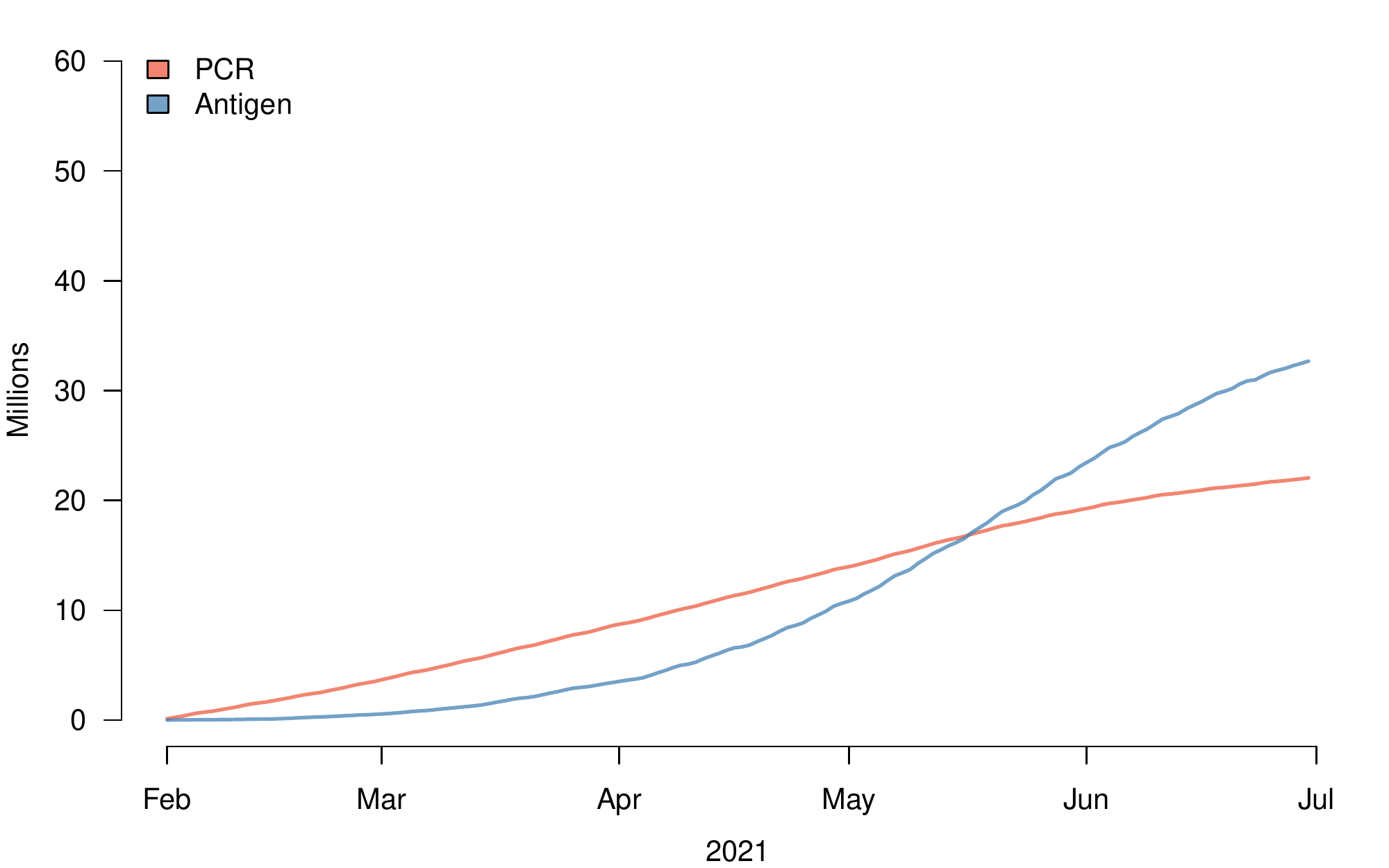}
	\caption{Cumulative number of tests recorded in Denmark between 1st February and 30th June 2021. The number of antigen tests (in blue) is initially low, increasing as they are rolled out nationwide during the 2nd quarter of 2021. PCR sampling (in red) has been used at a steady rate throughout, however by June 2021, the intensity of both test types begins to decrease.}\label{fig: cumulative test numbers}
\end{figure}

Data were collected from the Danish COVID-19 surveillance system based on a number of Danish national database/registries as follows:  date/time of all PCR samples and antigen tests in Denmark was taken from MiBa; age, sex and home address for all individuals registered in Denmark (approximately 6M people) was taken from the centralised person registry (CPR); vaccination status of all individuals registered in Denmark was taken from Det Danske Vaccinationsregister (DDV). Encrypted personal identifiers were used to link each individual test result to the person's sex, age, parish of residence and vaccine status. The CPR registry is predominantly static, although changes do occur due to relocation (including in and out of the country), births and deaths. For this study, we used a snapshot from 30th June 2021 as a reference for the whole population. Since antigen tests were not used extensively in Denmark until February 2021, we restricted the study period to between 1st February 2021 and 30th June 2021. During this period, the Alpha variant dominated the pandemic in Denmark, with the Delta variant becoming the principal variant in July 2021, i.e. after the study period. Approximately 22.1M (22,052,829) PCR and 32.7M (32,789,084) antigen tests were administered during the study period in total (Figure \ref{fig: cumulative test numbers}). This includes tests taken on a voluntary basis, as well as standard-procedure screening tests in hospitals. Since we do not know the reason that individual tests were taken, we must assume a mixed population among these test results. This group therefore includes both individuals with and without symptoms and individuals in need of a test result in order to attend social gatherings, visits, work, etc. Although various antigen test kits have been used in Denmark (see Table \ref{tab: antigen clinical performance}), the MiBa database does not contain information on which antigen test kit was used for a given test and as such, we cannot assess the performance of the kits individually. We therefore refer to antigen tests in general and the estimated performance is consequently the overall average performance of the different kits used in Denmark. Similarly, the MiBa database contains results from a number of different Nucleic Acid Amplification Test (NAAT) procedures other than RT-PCR, but we were unable to distinguish these based on the data available. However, the vast majority of these samples were analysed using RT-PCR during the study period, so we refer to all NAAT tests as `PCR' for the purposes of simplification.


\paragraph*{Data subset used for modelling}

A fundamental requirement of LCM is that two contemporaneous tests are available from the same individual, i.e. that paired test results are available. An equally important assumption is that the two test results are conditionally independent, i.e. that the decision to undertake one of the tests was not made conditional on the result of the other test, as may be the case for a confirmatory PCR test following a positive antigen test. We therefore restrict the case definition for paired observations used for the model to individuals where the PCR sampling preceded the antigen test by no more than 10 hours. We assume that this results in conditionally independent data based on the following reasoning:

\begin{itemize}
    \item The usual time for obtaining a PCR test result in Denmark is more than 10 hours, so in almost all cases, the PCR result would not be known at the time of the antigen test.
    \item Under Danish regulations in force during the relevant time period, a positive antigen test result was considered to be `overruled' by a negative PCR test due to the high specificity of the latter and relatively large number of antigen tests being performed. However, the converse was not true, i.e. a negative antigen test following a PCR could not be used by the individual to avoid isolation. It is therefore highly unlikely that an individual would take an antigen test if already in possession of a recent PCR test result.
\end{itemize}


\paragraph*{Data processing}

We identified the valid pairs of PCR samples and antigen tests where the antigen test followed the PCR sample by no more than 10 hours. This was done using the timestamps identifying when each test sample was collected, as registered in the MiBa database. We allow the same individual to appear multiple times in the data if they possessed multiple pairs of PCR$\to$antigen test results, except if the subsequent PCR sample was within 2 weeks of the previous PCR sample. We refer to this subset of test results as the `model data'.

In order to use LCM to analyse the data, it was necessary to stratify the data into multiple populations with varying prevalence. For the purposes of this study, these populations were generated artificially in order to maximise the statistical power of the LCM. We note that the artificial nature of these populations renders the estimates of prevalence effectively meaningless, but we can assume that the estimates of sensitivity and specificity are unbiased relative to those that would be obtained from a completely random sample. The procedure used for tests performed on unvaccinated individuals was as follows:

\begin{itemize}
    \item Each of the 22.1M PCR samples taken during the study period was assigned to one of 2,157 parishes (Danish `sogne') based on the registered home address of the individual being tested.
    \item PCR samples that were also included in the model data subset (see above) were removed.
    \item The proportion of the remaining PCR samples corresponding to a positive result was calculated per parish.
    \item The parishes were split into low-, medium- and high-prevalence groups based on this observed proportion.
    \item The two cutoff points were set so that the total population of each group of parishes (low, medium, high prevalence) was balanced to give approximately the same sample size.
    \item Each test pair within the model data was then linked to the parish of the individual being tested, and subsequently to the low-, medium- or high-prevalence group via the parish. 
\end{itemize}
    
\begin{figure}[ht]
	\centering
	\includegraphics[width=.7\textwidth]{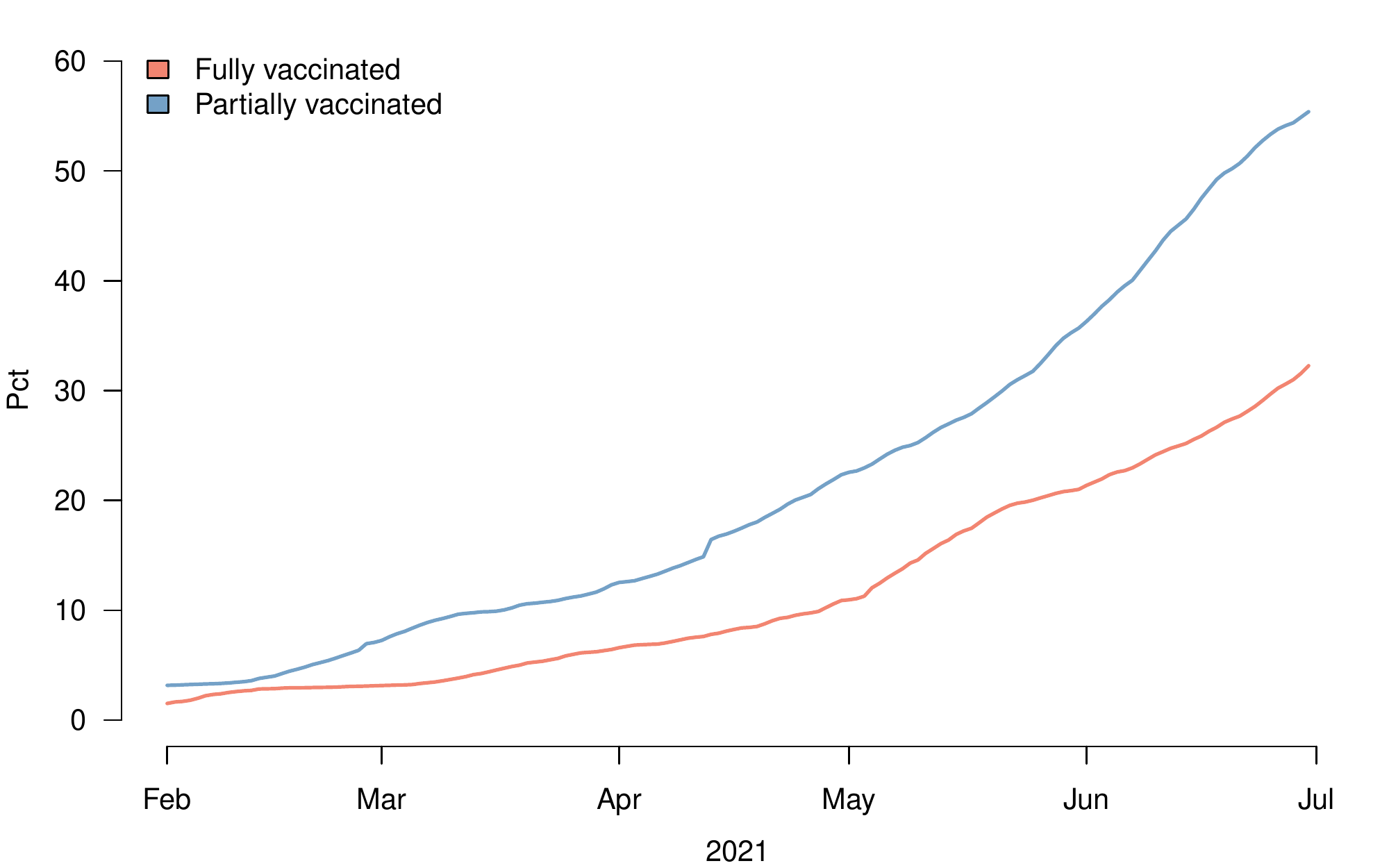}
	\caption{Vaccine coverage of fully vaccinated (red) and partially vaccinated (blue) individuals in Denmark between 1st February and 30th June 2021. Partially vaccinated means that the first dose has been administered.}\label{fig: vaccination status}
\end{figure}

Alongside these three prevalence-based groups for tests performed on unvaccinated individuals, a fourth group was established consisting of tests carried out on all partially and fully vaccinated individuals. The vaccination group was added to investigate whether any differences in sensitivity and specificity could be detected between vaccinated and unvaccinated individuals. Vaccines in Denmark have been administered in risk-based groups that are heavily correlated with age. Figure \ref{fig: vaccination status} shows the vaccine roll-out among the full population during the study period. As of 30th June 2021, 55.4\% of the population were at least partially vaccinated and 32.3\% were fully vaccinated. The two mRNA vaccines Comirnaty (Pfizer-BioNTech) and Spikevax (Moderna) have primarily been rolled out in Denmark as both Vaxzevria (AstraZeneca) and the COVID-19 vaccine Janssen (Johnson \& Johnson) ceased being used in April and May, respectively, on suspicion of severe side effects.  


\paragraph*{Statistical Modelling}

We fit a modified version of a standard two-test, four-population LCM to the paired test data obtained as described above. We follow the standard assumption of consistent specificity of both tests across all four populations. However, we allow the sensitivity of antigen testing and PCR sampling to vary between unvaccinated and vaccinated (including partly vaccinated) individuals, in order to allow for the possibility of reduced sensitivity due to suppression of viral excretion conditional on vaccination status. The sensitivity of both tests was assumed to be constant across the three unvaccinated populations.

We fit the LCM within a Bayesian framework, which requires prior distributions to be specified for all parameters. Minimally informative $Beta(1,1)$ priors were used for the prevalence parameters corresponding to each of the four groups, and weakly informative $Beta(2,1)$ priors were used for the sensitivity and specificity of both tests (two specificity and four sensitivity parameters). The model was fit using Markov chain Monte Carlo (MCMC) methods implemented using JAGS \cite{plummer2003jags}, interfaced from R \cite{team2020r} using the runjags package \cite{denwood2016runjags}. A burn-in period of 10,000 iterations was used before sampling 50,000 iterations from the posterior of each of two parallel chains. Convergence diagnostics were assessed using the Gelman-Rubin statistic and visual examination of trace plots \cite{gelman1992inference,toft2007assessing}, and the effective sample size of all parameters was checked to ensure that it exceeded 1,000 independent samples. The R code needed to replicate the Hui-Walter model discussed is provided as supplementary material. 

A sensitivity analysis was also performed in order to assess the impact of the 10-hour cutoff between PCR$\to$antigen test as described above. The data were re-tabulated for time lag values ranging between 1 and 24 hours, and the model was re-run using each of these datasets as input.

\paragraph*{Serial testing scheme}

The standard testing procedure in Denmark over the period February to July 2021 was to recommend the use of a confirmatory PCR test following a positive antigen result, in order to reduce false positive test results due to the presumed lower specificity of antigen tests. This serial scheme means that a negative antigen result does not require a confirmatory PCR, implying that the serial scheme will have a higher specificity but lower sensitivity than antigen testing alone. Using estimates of sensitivity and specificity of the two tests, we can derive the sensitivity and specificity for the serial scheme where both tests must be positive for the final result to be labelled positive, as
\begin{align*}
    \text{se}_\text{serial} &= \text{se}_\text{antigen}\cdot\text{se}_\text{PCR} \\
    \text{sp}_\text{serial} &= 1-(1-\text{sp}_\text{antigen})(1-\text{sp}_\text{PCR}).
\end{align*}
This implies that the serial sensitivity will decrease, whereas the specificity will increase, compared to either of the antigen or PCR tests used alone. For a population of size $N$ and prevalence $p$, this will result in fewer false positives: $N(1-p)(1-\text{sp})$, at a cost of more false negatives: $Np(1-\text{se})$, where the trade-off is balanced by $p$, se and sp. 

To illustrate the results, we calculated the  total number of false positives and false negatives expected under serial testing compared to antigen testing alone, assuming a known true prevalence. This procedure used Monte Carlo integration based on the estimates from each iteration of the LCM in order to obtain a full posterior distribution for all parameter estimates. Results are presented as expected cases per 10,000, as well as the expected cases in a population size of 32,789,084, which corresponds to the number of antigen tests in Denmark during the study period. In order to assess the impact of prevalence, we calculated the false positives and false negatives for prevalence values ranging between 0.01\% and 4\%.

\section*{Ethical statement}
This work was carried out exclusively using existing data contained within the Danish national data registries. Individual-level data were processed within the secure computing environment provided by the Danish Health Data Authority and were aggregated before being extracted in the form of the results presented, which do not contain individual-level data of any kind.  Therefore, this study does not require ethical approval under Danish law or consideration under the European General Data Protection Regulation (GDPR).

\section*{Acknowledgements}
The authors declare that they have no competing interests. All data as well as R code needed to evaluate the conclusions in the paper are provided in the paper and/or the Supplementary Materials.  The authors received funding from the Danish government (as part of SSI's Expert Modelling Group) concurrently with undertaking this research.
Author contributions: JSØ processed and analysed the data and was a principle contributor to writing the text. CK contributed to the formulation of ideas and was a principle contributor to writing the text. LEC contributed to the formulation of ideas, critical discussion of assumptions in the model, and review and editing of the text. MAA, CHM \& MVL contributed specific sections of the text and edited the entire text. MD contributed to processing and analysing the data, developed the LCM used for the analysis, and was a principle contributor to writing the text. All authors reviewed and approved the final version of the manuscript.  We would like to thank Sarah Layhe (Vaetta Editing) for proofreading the manuscript.


%

\bibliography{scibib}

\bibliographystyle{Science}


\newpage
\section*{Supplementary materials}

\begin{figure}[ht]
	\centering
	\includegraphics[width=.7\textwidth]{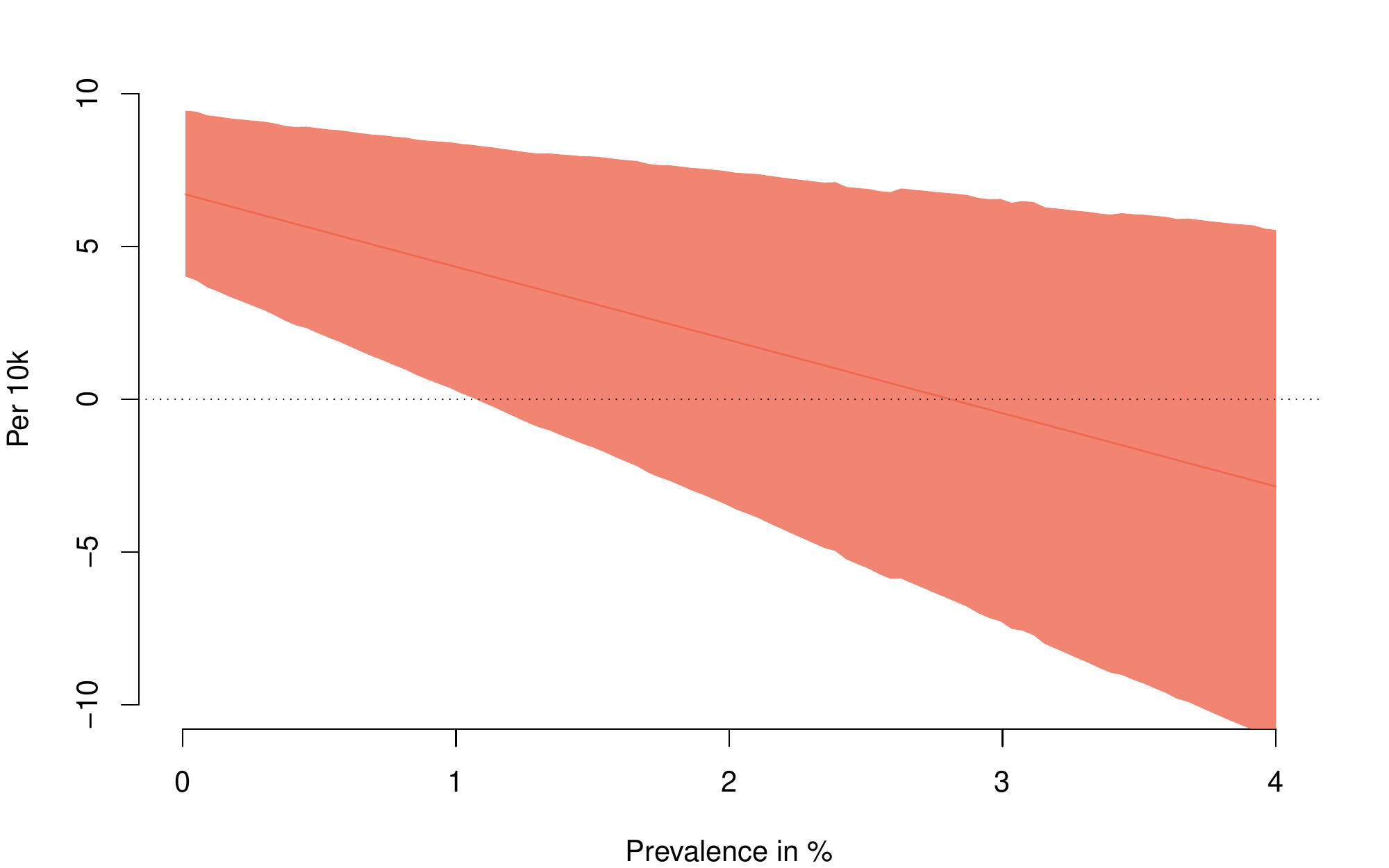}
	\caption{The estimated balance between the decrease in false positive cases and increase in false negative cases per 10,000 individuals, with 95\% confidence limits, as a result of changing from an antigen test alone to a serial testing scheme. The prevalence varies from 0 to 4\% and shows that as the prevalence increases, removing false positives by employing a serial testing scheme will cost more in terms of increasing false negative cases. The estimated median shows that the balance tips at a prevalence of 3\%, such that more false negative cases can be expected than the number of removed false positive cases.}\label{fig: fn cases}
\end{figure}


\begin{table}[ht]
\centering
\begin{tabular}{lll r}
\toprule
Sub-population & PCR result & Antigen result & Tally \\
\midrule
    High prevalence & Positive & Positive & 1,030 \\ 
    High prevalence & Positive & Negative & 988 \\ 
    High prevalence & Negative & Positive & 83 \\ 
    High prevalence & Negative & Negative & 75,338 \\ 
\midrule
Total        &          &          & 77,439 \\
\midrule
\midrule
    Medium prevalence & Positive & Positive & 557 \\ 
    Medium prevalence & Positive & Negative & 585 \\ 
    Medium prevalence & Negative & Positive & 83 \\ 
    Medium prevalence & Negative & Negative & 61,612 \\ 
\midrule
Total        &          &          & 62,837 \\
\midrule
\midrule
    Low prevalence & Positive & Positive & 269 \\ 
    Low prevalence & Positive & Negative & 311 \\ 
    Low prevalence & Negative & Positive & 52 \\ 
    Low prevalence & Negative & Negative & 57,360 \\ 
\midrule
Total        &          &          & 57,992 \\
\midrule    
\midrule    
    Vaccinated & Positive & Positive & 151 \\ 
    Vaccinated & Positive & Negative & 180 \\ 
    Vaccinated & Negative & Positive & 27 \\ 
    Vaccinated & Negative & Negative & 40,595 \\ 
\midrule
Total        &          &          & 40,953 \\
\midrule    
\midrule
Grand total  &          &          & 239,221 \\
\bottomrule
\end{tabular}
\caption{Frequencies for the PCR$\to$antigen test data between 1st February 2021 and 30th June 2021 for a 10-hour delay between tests. The total of 239,221 pairs from 222,805 unique individuals were extracted from a pool of $\approx$55M tests. The population of Denmark is $\approx$6M.}\label{tab: RT-PCR -> antigen}
\end{table}

\begin{table}[ht]
\centering
\begin{tabular}{ll rrr}
\toprule
Test & Parameter & Median & 2.5\% & 97.5\% \\
\midrule
\multirow{3}{*}{Antigen} & Se               & 53.82 & 49.83 & 57.93 \\ 
                         & Se (vaccinated)  & 56.01 & 44.50 & 69.84 \\ 
                         & Sp               & 99.93 & 99.91 & 99.96 \\ 
\midrule                         
\multirow{3}{*}{PCR}     & Se               & 95.68 & 92.79 & 98.43 \\ 
                         & Se (vaccinated)  & 97.44 & 91.55 & 100.00 \\ 
                         & Sp               & 99.85 & 99.73 & 99.97 \\ 
\midrule
\multirow{3}{*}{Serial}  & Se               & 51.48 & 47.37 & 55.96 \\ 
                         & Se (vaccinated)  & 54.22 & 42.69 & 68.35 \\ 
                         & Sp               & 100.00 & 100.00 & 100.00 \\ 
\bottomrule
\end{tabular}
\caption{LCM estimates (in \%) from the model data. The median estimates show that the specificity is close to 100\% for all types of tests. Sensitivities are estimated for the low/medium/high groups combined and the vaccinated group separately. These are around 96-97\% for PCR and 54-56\% for antigen tests for a 10-hour delay. The serial testing scheme assumes that a PCR sample is taken as a follow-up to a positive antigen test, with the overall result considered positive if both tests are positive. The 100.00\% figures for specificity are due to rounding, but evidently suggest a near-perfect specificity for serial testing.}\label{tab: HW estimates RT-PCR -> antigen}
\end{table}

\begin{table}[ht]
\centering
\begin{tabular}{lr | rrr | rrr}
    \toprule
  & & \multicolumn{3}{c}{Cases in Denmark} & \multicolumn{3}{c}{Cases per 10,000} \\
Type & $p$ & Median & 2.5\% & 97.5\% & Median & 2.5\% & 97.5\% \\ 
    \midrule
    Antigen FP & 0.01\% & 22,098 & 13,311 & 31,015 & 6.74 & 4.07 & 9.47 \\ 
    Antigen FP & 0.1\% & 22,078 & 13,332 & 31,020 & 6.73 & 4.09 & 9.48 \\ 
    Antigen FP & 0.4\% & 22,011 & 13,259 & 30,894 & 6.71 & 4.05 & 9.43 \\ 
    Antigen FP & 1\% & 21,879 & 13,179 & 30,708 & 6.67 & 4.02 & 9.37 \\ 
    Antigen FP & 2\% & 21,658 & 13,046 & 30,398 & 6.61 & 3.99 & 9.28 \\ 
    Antigen FP & 3\% & 21,437 & 12,913 & 30,088 & 6.54 & 3.94 & 9.18 \\ 
    Antigen FP & 4\% & 21,216 & 12,780 & 29,778 & 6.47 & 3.9 & 9.08 \\ 
    \midrule
    Serial FP & 0.01\% & 32 & 3 & 65 & 0.01 & 0 & 0.02 \\ 
    Serial FP & 0.1\% & 32 & 3 & 65 & 0.01 & 0 & 0.02 \\ 
    Serial FP & 0.4\% & 32 & 3 & 65 & 0.01 & 0.00 & 0.02 \\ 
    Serial FP & 1\% & 32 & 3 & 64 & 0.01 & 0 & 0.02 \\ 
    Serial FP & 2\% & 31 & 3 & 64 & 0.01 & 0 & 0.02 \\ 
    Serial FP & 3\% & 31 & 3 & 63 & 0.01 & 0 & 0.02 \\ 
    Serial FP & 4\% & 31 & 3 & 62 & 0.01 & 0 & 0.02 \\ 
    \midrule    
    Antigen FN & 0.01\% &   1,514 &   1,379 &   1,645 & 0.46 & 0.42 & 0.5 \\ 
    Antigen FN & 0.1\% &  15,141 &  13,794 &  16,450 & 4.62 & 4.21 & 5.02 \\ 
    Antigen FN & 0.4\% & 60,564 & 55,177 & 65,800 & 18.47 & 16.83 & 20.07 \\  
    Antigen FN & 1\% & 151,410 & 137,941 & 164,499 & 46.18 & 42.07 & 50.17 \\ 
    Antigen FN & 2\% & 302,820 & 275,883 & 328,998 & 92.35 & 84.14 & 100.34 \\ 
    Antigen FN & 3\% & 454,230 & 413,824 & 493,497 & 138.53 & 126.21 & 150.51 \\ 
    Antigen FN & 4\% & 605,640 & 551,766 & 657,996 & 184.71 & 168.28 & 200.68 \\ 
    \midrule
    Serial FN & 0.01\% &   1,591 &   1,444 &   1,726 & 0.49 & 0.44 & 0.53 \\ 
    Serial FN & 0.1\% &  15,909 &  14,439 &  17,258 & 4.85 & 4.4 & 5.26 \\ 
    Serial FN & 0.4\% & 63,635 & 57,756 & 69,033 & 19.41 & 17.61 & 21.05 \\  
    Serial FN & 1\% & 159,088 & 144,391 & 172,582 & 48.52 & 44.04 & 52.63 \\ 
    Serial FN & 2\% & 318,175 & 288,782 & 345,163 & 97.04 & 88.07 & 105.27 \\ 
    Serial FN & 3\% & 477,263 & 433,174 & 517,745 & 145.56 & 132.11 & 157.9 \\ 
    Serial FN & 4\% & 636,351 & 577,565 & 690,326 & 194.07 & 176.15 & 210.54 \\ 
  \bottomrule
\end{tabular}
\caption{Estimated false positive (FP) and false negative (FN) cases for 32,789,084 tests and per 10,000 individuals for both antigen testing alone and serial (antigen $\to$ PCR) testing for varying prevalence $p$. Serial testing essentially reduces the false positive cases to near zero while increasing the false negative rates.}
\label{tab: estimate fp and fn cases for antigen and serial testing}
\end{table}

\begin{table}[ht]
\centering
\begin{tabular}{lr | rrr | rrr}
    \toprule
  & & \multicolumn{3}{c}{Cases in Denmark} & \multicolumn{3}{c}{Cases per 10,000} \\
Type & $p$ & Median & 2.5\% & 97.5\% & Median & 2.5\% & 97.5\% \\ 
    \midrule
    FN increase & 0.01\% &     76 &     26 &    125 & 0.02 & 0.01 & 0.04 \\ 
    FN increase & 0.1\% &    763 &    263 &  1,253 & 0.23 & 0.08 & 0.38 \\ 
    FN increase & 0.4\% & 3,054 & 1,050 & 5,012 & 0.93 & 0.32 & 1.53 \\
    FN increase & 1\% &  7,634 &  2,626 & 12,530 & 2.33 & 0.8 & 3.82 \\ 
    FN increase & 2\% & 15,269 &  5,252 & 25,059 & 4.66 & 1.6 & 7.64 \\ 
    FN increase & 3\% & 22,903 &  7,878 & 37,589 & 6.99 & 2.4 & 11.46 \\ 
    FN increase & 4\% & 30,538 & 10,504 & 50,118 & 9.31 & 3.2 & 15.29 \\ 
    \midrule
    FP decrease & 0.01\% & 22,076 & 13,378 & 31,068 & 6.73 & 4.08 & 9.48 \\ 
    FP decrease & 0.1\% & 22,056 & 13,366 & 31,040 & 6.73 & 4.08 & 9.47 \\
    FP decrease & 0.4\% & 21,990 & 13,326 & 30,947 & 6.71 & 4.06 & 9.44 \\ 
    FP decrease & 1\% & 21,857 & 13,246 & 30,761 & 6.67 & 4.04 & 9.38 \\ 
    FP decrease & 2\% & 21,636 & 13,112 & 30,450 & 6.6 & 4 & 9.29 \\ 
    FP decrease & 3\% & 21,416 & 12,978 & 30,139 & 6.53 & 3.96 & 9.19 \\ 
    FP decrease & 4\% & 21,195 & 12,844 & 29,829 & 6.46 & 3.92 & 9.1 \\ 
    \midrule
    FP/FN balance & 0.01\% & 22,000 &  13,198 & 30,974 & 6.71 & 4.02 & 9.45 \\ 
    FP/FN balance & 0.1\% & 21,292 &  11,906 & 30,447 & 6.49 & 3.63 & 9.29 \\
    FP/FN balance & 0.4\% & 18,930 & 8,121 & 29,267 & 5.77 & 2.48 & 8.93 \\ 
    FP/FN balance & 1\% & 14,222 &     870 & 27,455 & 4.34 & 0.27 & 8.37 \\ 
    FP/FN balance & 2\% &  6,361 & -11,432 & 24,472 & 1.94 & -3.49 & 7.46 \\ 
    FP/FN balance & 3\% & -1,494 & -23,934 & 21,493 & -0.46 & -7.3 & 6.56 \\ 
    FP/FN balance & 4\% & -9,361 & -36,794 & 18,175 & -2.86 & -11.22 & 5.54 \\ 
  \hline
\end{tabular}
\caption{Estimated increases in false negative (FN) and false positive (FP) cases for 32,789,084 tests and per 10,000 individuals for varying prevalence $p$. In addition, the table also shows the estimated balance between false positives and false negatives. When the prevalence $p$ reaches $\approx$3\%, the median increase in false negatives balances the false positives. At the lower limit of the confidence interval (2.5\%), this balance occurs slightly above $p=1\%$.}
\label{tab: fp fn difference estimates}
\end{table}

\clearpage

\appendix
\section{Additional results}\label{app: additional results}

\begin{table}[ht]
\centering
\begin{tabular}{rrrr | rrr}
  \toprule
  & \multicolumn{3}{c}{Antigen} & \multicolumn{3}{c}{PCR} \\
Lag (hours) & Median & 2.5\% & 97.5\% & Median & 2.5\% & 97.5\% \\ 
  \midrule
  1 & 53.60 & 49.54 & 58.37 & 96.97 & 94.19 & 99.87 \\ 
    2 & 54.06 & 49.83 & 58.32 & 97.65 & 95.07 & 100.00 \\ 
    3 & 54.44 & 50.13 & 58.64 & 97.49 & 94.99 & 99.99 \\ 
    4 & 54.59 & 50.39 & 59.05 & 97.35 & 94.91 & 99.97 \\ 
    5 & 53.03 & 49.38 & 57.13 & 96.77 & 94.01 & 99.60 \\ 
    6 & 52.67 & 48.96 & 56.61 & 96.83 & 94.18 & 99.56 \\ 
    7 & 53.33 & 49.57 & 57.45 & 96.32 & 93.58 & 99.08 \\ 
    8 & 53.36 & 49.61 & 57.42 & 96.07 & 93.26 & 98.94 \\ 
    9 & 53.49 & 49.54 & 57.51 & 95.93 & 93.13 & 98.74 \\ 
  10 & 53.82 & 49.83 & 57.95 & 95.68 & 92.87 & 98.51 \\ 
  11 & 53.72 & 49.89 & 57.72 & 95.88 & 93.06 & 98.68 \\ 
  12 & 54.03 & 50.04 & 58.14 & 95.58 & 92.63 & 98.45 \\ 
  13 & 54.28 & 50.39 & 58.56 & 95.55 & 92.72 & 98.57 \\ 
  14 & 54.45 & 50.48 & 58.64 & 95.78 & 92.74 & 98.57 \\ 
  15 & 54.38 & 50.29 & 58.37 & 95.82 & 92.81 & 98.65 \\ 
  16 & 54.67 & 50.61 & 58.97 & 95.98 & 92.83 & 98.84 \\ 
  17 & 54.84 & 50.72 & 58.97 & 95.71 & 92.70 & 98.75 \\ 
  18 & 55.00 & 51.07 & 59.27 & 95.59 & 92.51 & 98.49 \\ 
  19 & 54.90 & 50.90 & 58.96 & 95.54 & 92.36 & 98.44 \\ 
  20 & 54.54 & 50.61 & 58.39 & 96.01 & 92.97 & 98.98 \\ 
  21 & 54.56 & 50.74 & 58.31 & 95.89 & 92.80 & 98.90 \\ 
  22 & 54.50 & 50.81 & 58.30 & 95.66 & 92.56 & 98.60 \\ 
  23 & 54.41 & 50.86 & 58.09 & 95.50 & 92.29 & 98.45 \\ 
  24 & 54.49 & 50.85 & 58.07 & 94.66 & 91.62 & 97.69 \\ 
  \bottomrule
\end{tabular}
\caption{LCM estimates of sensitivity (in \%) for the low-, medium- and high-prevalence parish groups for varying time lags between tests. Confidence intervals for both tests overlap across all time lags, thus indicating no trend in test performance up to a lag of 24 hours.}
\end{table}

\begin{table}[ht]
\centering
\begin{tabular}{rrrr | rrr}
  \toprule
  & \multicolumn{3}{c}{Antigen} & \multicolumn{3}{c}{PCR} \\
Lag (hours) & Median & 2.5\% & 97.5\% & Median & 2.5\% & 97.5\% \\ 
  \midrule
    1 & 99.93 & 99.89 & 99.97 & 99.84 & 99.68 & 100.00 \\ 
    2 & 99.91 & 99.88 & 99.94 & 99.83 & 99.68 & 99.99 \\ 
    3 & 99.91 & 99.88 & 99.94 & 99.82 & 99.68 & 99.98 \\ 
    4 & 99.92 & 99.89 & 99.95 & 99.82 & 99.68 & 99.97 \\ 
    5 & 99.93 & 99.90 & 99.95 & 99.87 & 99.75 & 100.00 \\ 
    6 & 99.92 & 99.90 & 99.95 & 99.87 & 99.76 & 100.00 \\ 
    7 & 99.93 & 99.90 & 99.96 & 99.86 & 99.74 & 99.99 \\ 
    8 & 99.93 & 99.90 & 99.96 & 99.86 & 99.74 & 99.99 \\ 
    9 & 99.93 & 99.90 & 99.96 & 99.86 & 99.74 & 99.98 \\ 
  10 & 99.93 & 99.90 & 99.96 & 99.85 & 99.73 & 99.97 \\ 
  11 & 99.93 & 99.90 & 99.96 & 99.86 & 99.74 & 99.98 \\ 
  12 & 99.93 & 99.90 & 99.96 & 99.86 & 99.74 & 99.98 \\ 
  13 & 99.93 & 99.90 & 99.96 & 99.85 & 99.74 & 99.98 \\ 
  14 & 99.93 & 99.90 & 99.96 & 99.85 & 99.73 & 99.97 \\ 
  15 & 99.93 & 99.90 & 99.96 & 99.86 & 99.75 & 99.98 \\ 
  16 & 99.93 & 99.90 & 99.96 & 99.85 & 99.73 & 99.96 \\ 
  17 & 99.93 & 99.90 & 99.96 & 99.85 & 99.74 & 99.96 \\ 
  18 & 99.93 & 99.91 & 99.96 & 99.85 & 99.75 & 99.96 \\ 
  19 & 99.93 & 99.91 & 99.96 & 99.86 & 99.76 & 99.96 \\ 
  20 & 99.93 & 99.90 & 99.95 & 99.87 & 99.78 & 99.97 \\ 
  21 & 99.93 & 99.90 & 99.95 & 99.88 & 99.79 & 99.97 \\ 
  22 & 99.93 & 99.90 & 99.95 & 99.88 & 99.80 & 99.97 \\ 
  23 & 99.92 & 99.90 & 99.95 & 99.89 & 99.81 & 99.97 \\ 
  24 & 99.93 & 99.91 & 99.95 & 99.89 & 99.82 & 99.97 \\    \bottomrule
\end{tabular}
\caption{LCM estimates of specificity (in \%) for the low-, medium- and high-prevalence parish groups for varying time lags between tests. Confidence intervals for both tests overlap across all time lags, thus indicating no trend in test performance up to a lag of 24 hours.}
\end{table}

\begin{landscape} 

\begin{table}[ht]
\centering
\begin{tabular}{rrrr | rrr | rrr | rrr}
  \toprule
  & \multicolumn{3}{c}{High} & \multicolumn{3}{c}{Medium} & \multicolumn{3}{c}{Low} & \multicolumn{3}{c}{Vaccinated} \\
Lag (hours) & Median & 2.5\% & 97.5\% & Median & 2.5\% & 97.5\% & Median & 2.5\% & 97.5\% & Median & 2.5\% & 97.5\% \\ 
  \midrule
  1 & 3.15 & 2.88 & 3.41 & 2.25 & 1.99 & 2.50 & 1.11 & 0.90 & 1.31 & 0.97 & 0.73 & 1.19 \\ 
    2 & 3.01 & 2.78 & 3.25 & 2.12 & 1.90 & 2.35 & 1.06 & 0.88 & 1.25 & 0.90 & 0.67 & 1.10 \\ 
    3 & 2.90 & 2.68 & 3.13 & 2.04 & 1.82 & 2.26 & 1.01 & 0.84 & 1.18 & 0.82 & 0.63 & 1.02 \\ 
    4 & 2.80 & 2.58 & 3.01 & 1.94 & 1.73 & 2.15 & 0.97 & 0.81 & 1.14 & 0.78 & 0.59 & 0.97 \\ 
    5 & 2.79 & 2.59 & 2.99 & 1.92 & 1.73 & 2.12 & 0.98 & 0.82 & 1.14 & 0.80 & 0.62 & 0.97 \\ 
    6 & 2.73 & 2.53 & 2.92 & 1.88 & 1.69 & 2.07 & 0.96 & 0.82 & 1.12 & 0.77 & 0.60 & 0.93 \\ 
    7 & 2.68 & 2.48 & 2.87 & 1.84 & 1.65 & 2.03 & 0.93 & 0.78 & 1.08 & 0.73 & 0.56 & 0.89 \\ 
    8 & 2.64 & 2.44 & 2.82 & 1.82 & 1.63 & 2.00 & 0.92 & 0.78 & 1.07 & 0.72 & 0.55 & 0.87 \\ 
    9 & 2.59 & 2.41 & 2.78 & 1.80 & 1.61 & 1.98 & 0.91 & 0.77 & 1.07 & 0.70 & 0.54 & 0.86 \\ 
  10 & 2.56 & 2.38 & 2.76 & 1.75 & 1.58 & 1.94 & 0.89 & 0.75 & 1.04 & 0.68 & 0.52 & 0.84 \\ 
  11 & 2.54 & 2.35 & 2.72 & 1.75 & 1.57 & 1.92 & 0.90 & 0.76 & 1.04 & 0.68 & 0.52 & 0.83 \\ 
  12 & 2.50 & 2.32 & 2.69 & 1.73 & 1.56 & 1.91 & 0.89 & 0.75 & 1.04 & 0.67 & 0.52 & 0.84 \\ 
  13 & 2.48 & 2.30 & 2.66 & 1.71 & 1.54 & 1.89 & 0.88 & 0.75 & 1.03 & 0.67 & 0.52 & 0.83 \\ 
  14 & 2.43 & 2.25 & 2.61 & 1.68 & 1.51 & 1.85 & 0.86 & 0.73 & 1.01 & 0.67 & 0.52 & 0.83 \\ 
  15 & 2.38 & 2.21 & 2.57 & 1.65 & 1.48 & 1.81 & 0.85 & 0.72 & 0.99 & 0.66 & 0.51 & 0.82 \\ 
  16 & 2.32 & 2.15 & 2.49 & 1.59 & 1.42 & 1.75 & 0.82 & 0.69 & 0.96 & 0.65 & 0.50 & 0.81 \\ 
  17 & 2.26 & 2.09 & 2.43 & 1.53 & 1.37 & 1.69 & 0.79 & 0.66 & 0.92 & 0.65 & 0.50 & 0.80 \\ 
  18 & 2.20 & 2.05 & 2.37 & 1.48 & 1.33 & 1.63 & 0.76 & 0.64 & 0.89 & 0.63 & 0.49 & 0.78 \\ 
  19 & 2.14 & 1.98 & 2.29 & 1.42 & 1.28 & 1.57 & 0.73 & 0.62 & 0.85 & 0.62 & 0.48 & 0.76 \\ 
  20 & 2.07 & 1.93 & 2.23 & 1.36 & 1.23 & 1.50 & 0.69 & 0.58 & 0.80 & 0.61 & 0.48 & 0.74 \\ 
  21 & 2.02 & 1.89 & 2.17 & 1.30 & 1.18 & 1.43 & 0.67 & 0.57 & 0.78 & 0.60 & 0.47 & 0.72 \\ 
  22 & 1.96 & 1.83 & 2.09 & 1.25 & 1.13 & 1.37 & 0.64 & 0.54 & 0.74 & 0.57 & 0.46 & 0.70 \\ 
  23 & 1.92 & 1.79 & 2.04 & 1.20 & 1.09 & 1.32 & 0.62 & 0.53 & 0.72 & 0.57 & 0.46 & 0.68 \\ 
  24 & 1.86 & 1.74 & 1.98 & 1.15 & 1.04 & 1.25 & 0.59 & 0.51 & 0.68 & 0.54 & 0.44 & 0.65 \\ 
  \bottomrule
\end{tabular}
\caption{LCM estimates of prevalence (in \%) for the four groups: low, medium and high prevalence as well as vaccinated, for varying time lags between tests.}
\end{table}
\end{landscape}

\pagebreak

\begin{landscape}
\begin{table}[ht]
\centering
\begin{tabular}{lllll}
\toprule
Manufacturer & Commercial Name & Sensitivity & Specificity & Remark \\
\midrule
Abbott Rapid Diagnostics & Panbio\texttrademark\enskip COVID-19 Ag Rapid Test  & 91.4\% & 99.8\% & NP swab (Ct $\leq33$) \\
                                                          && 98.1\% & 99.8\% & Nasal swab (Ct $\leq 33$)\\
\midrule
BIOSYNEX S.A. & BIOSYNEX COVID-19 Ag BSS & 96\% & 100\% & NP swab \\
\midrule
CTK Biotech, Inc & OnSite COVID-19 Ag Rapid Test & 92.3\% & 100\% & Nasal, NP swab\\
\midrule
Roche (SD BIOSENSOR) & SARS-CoV-2 Rapid antigen Test Nasal  & 89.6\% &  99.1\% & Nasal swab (Ct $\leq$ 30) \\ 
                                                          && 93.1\% & & Nasal swab (Ct $\leq$ 27) \\
\midrule
SD BIOSENSOR Inc. & STANDARD Q COVID-19 Ag Test & 96.52\% & 99.68\% & NP swab \\
\bottomrule
\end{tabular}
\caption{Reported clinical performance of antigen test kits used in Denmark. NP swab refers to nasopharyngeal swab. \\Source: \cite{EC2021}}\label{tab: antigen clinical performance}
\end{table}
\end{landscape}


\begin{table}[ht]
\centering
\begin{tabular}{l rrr}
\toprule
Group & Median & 2.5\% & 97.5\% \\
\midrule
High  & 2.56 & 2.37 & 2.75 \\ 
Medium  & 1.73 & 1.56 & 1.92 \\ 
Low  & 0.89 & 0.75 & 1.04 \\ 
Vaccinated & 0.68 & 0.52 & 0.84 \\
\bottomrule
\end{tabular}
\caption{LCM estimates of prevalence (in \%) based on the model data with a maximum 10-hour time lag between the tests.}\label{tab: HW prevalence RT-PCR -> antigen}
\end{table}

\begin{figure}[ht]
	\centering
	\includegraphics[width=.9\textwidth]{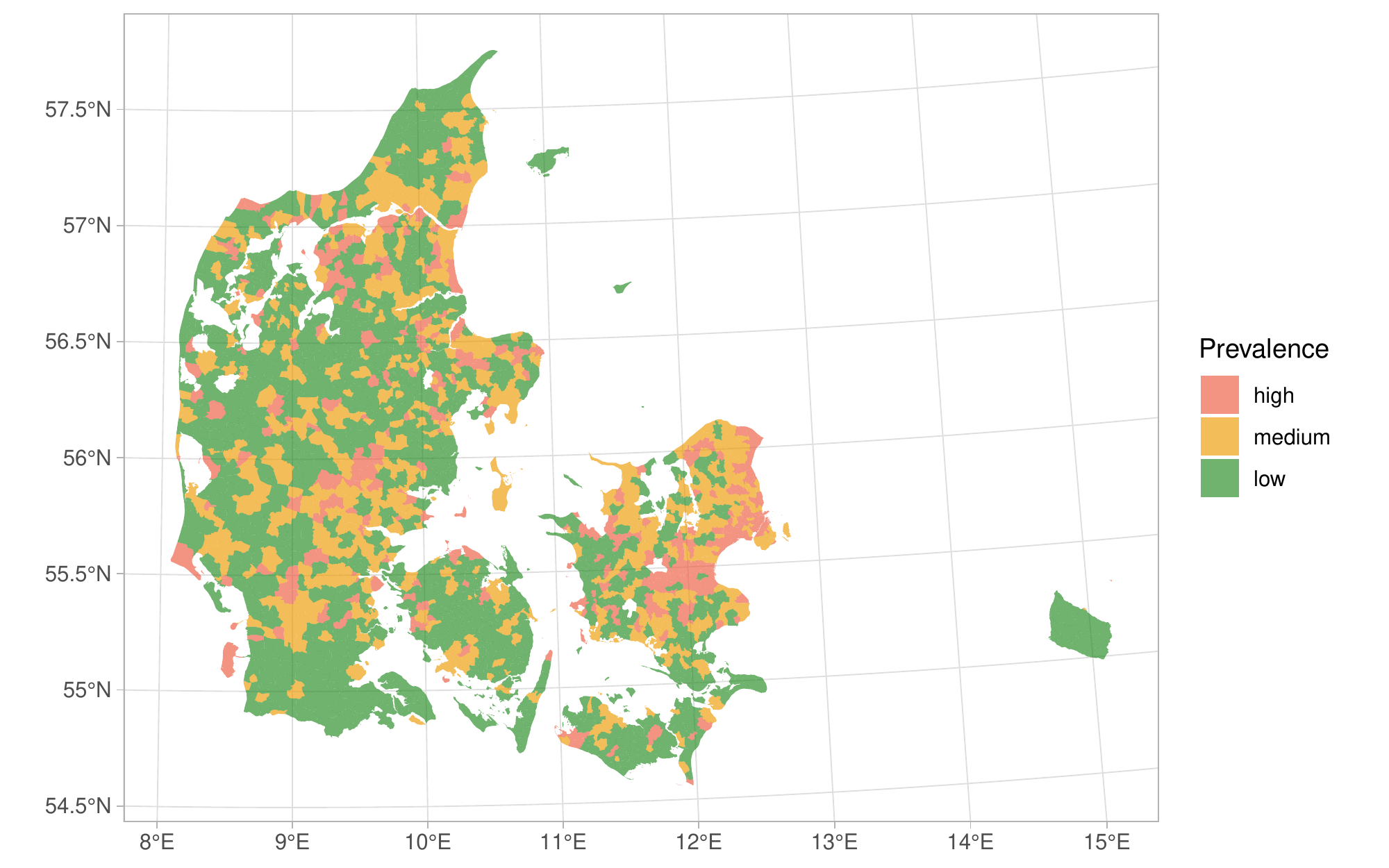}
	\caption{The geographical location of parishes coloured according to their prevalence range: low, medium and high. Even though there are clusters of high-prevalence parishes, these are not exclusive to densely populated areas and larger cities.}\label{fig: parish map}
\end{figure}

\end{document}